\documentclass[12pt,prl,aps,amssymb,amsmath,tightenlines]{revtex4} 

\usepackage[dvips]{graphicx}
\usepackage{dsfont}
\newcommand{\half}{\mbox{$\textstyle{\frac{1}{2}}$}}
\newcommand{\re}{{\rm e}}
\newcommand{\ri}{{\rm i}}
\newcommand{\rd}{{\rm d}}

\usepackage{xcolor}

\begin{document}
\title{Asymptotic analysis of a pseudo-Hermitian \\ Riemann-zeta Hamiltonian}
\author{Carl M. Bender$^1$ and Dorje C. Brody$^{2}$}

\affiliation{
$^1$Department of Physics, Washington University, St Louis, MO 63130, USA\\
$^2$
Department of Optical Physics and Modern Natural Science, St Petersburg
National Research University of Information Technologies, Mechanics and Optics, 
St Petersburg 197101, Russia}

\begin{abstract}
The differential-equation eigenvalue problem associated with a recently-introduced Hamiltonian, whose eigenvalues correspond to the zeros of the Riemann zeta function, is analyzed using Fourier and WKB analysis. The Fourier analysis leads to a challenging open problem concerning the formulation of the eigenvalue problem in the momentum space. The WKB analysis gives the exact asymptotic behavior of the eigenfunction. 
\end{abstract}

\maketitle

\noindent 
{\bf Introduction}. In a recent paper we introduced the Hamiltonian  
\begin{eqnarray}
{\hat H}=\frac{{\mathds 1}}{{\mathds 1}-\re^{-{\rm i}{\hat p}}}\left({\hat x}
{\hat p}+{\hat p}{\hat x}\right)({\mathds 1}-\re^{-{\rm i}{\hat p}}),
\label{e1}
\end{eqnarray}
which has the property that subject to the boundary condition $\psi(0)=0$ on its
eigenfunctions defined on the interval ${\mathds R}^+=(0,\infty)$, the 
eigenvalues $\{E_n\}$ of ${\hat H}$ are such that $\{\half(1-\ri E_n)\}$ are the
zeros of the Riemann zeta function \cite{r1}. The analysis of \cite{r1} requires
Borel summation to regulate a divergent sum. This divergent sum arises because
${\hat H}$ is an infinite-order differential operator in $x$-space. However, the
Hamiltonian ${\hat H}$ is a linear function of the position operator ${\hat x}$,
which suggests that in Fourier space the differential equation eigenvalue
problem ${\hat H}\psi(x)=E\psi(x)$ reduces to a much simpler linear first-order
differential equation, analogous to the eigenvalue problem associated with the
Berry-Keating Hamiltonian ${\hat h}_{\rm BK}={\hat x}{\hat p}+{\hat p}{\hat x}$
\cite{r4}, to which ${\hat H}$ is related by a similarity transformation.

This form of the eigenvalue problem motivates us to pursue the analysis in
Fourier space ($p$ space). However, as we will show, transforming the problem to
$p$ space is nontrivial and leads to difficulties. An alternative way to analyze
the infinite-order differential equation in $x$ space is to employ a WKB
approximation. Although the differential equation is of infinite order, it is
possible to calculate the terms in the WKB series in closed form to any order
in powers of $\hbar$. In this paper we calculate the first three terms in the
WKB series, which allows us to determine the asymptotic behavior of the
eigenfunctions for large $x$. The purpose of this note is (a) to explain the
difficulties encountered in the Fourier analysis, which lead to some interesting
open questions, and (b) to present our WKB analysis of the differential equation
eigenvalue problem ${\hat H}\psi(x)=E\psi(x)$, which gives the asymptotic
behavior of the eigenfunction $\psi(x)$. 

\medskip
\noindent{\bf Review of previous work}. We summarize briefly the relevant
findings of \cite{r1}. The Hamiltonian ${\hat H}$ is formally similar to the
Berry-Keating Hamiltonian ${\hat h}_{\rm BK}$ via the transformation ${\hat H}=
{\hat\Delta}^{-1}{\hat h}_{\rm BK}{\hat\Delta}$, where ${\hat\Delta}\equiv{
\mathds 1}-\re^{-{\rm i}{\hat p}}$ and ${\hat p}\equiv-\ri\rd/\rd x$. The action
of ${\hat\Delta}$ on a sufficiently smooth function $f(x)$ gives ${\hat\Delta}f(
x)=f(x)-f(x-1)$. If, in addition, $f(x)$ vanishes faster than $x^{-1}$ as $x\to
\infty$, then we have 
$${\hat\Delta}^{-1}f(x)=-\sum_{k=1}^\infty f(x+k).$$ 
This can be seen by writing $f(x+k)=\re^{{\rm i}k{\hat p}} f(x)$ and then
summing the geometric series. On the half line ${\mathds R}^+=(0,\infty)$ the
operator ${\hat p}$ is not selfadjoint \cite{refxx}, and since the imaginary
part of the spectrum of ${\hat p}$ on the space of square-integrable functions
on ${\mathds R}^+$ is strictly positive, the absolute value of the spectrum of
$\re^{{\rm i}{\hat p}}$ is less than one. Hence the geometric series above 
converges to ${\hat\Delta}^{-1}$. More generally, expanding ${\hat\Delta}^{-1}$
as a series in powers of ${\hat p}$, we obtain 
$$
{\hat\Delta}^{-1}=\frac{{\mathds 1}}{\ri{\hat p}}\,\sum_{n=0}^\infty B_n\frac{
(-\ri{\hat p})^n}{n!},
$$
where $\{B_k\}$ are the Bernoulli numbers (with the convention that $B_1=-
\frac{1}{2}$). The operator $(\ri{\hat p})^{-1}$ is interpreted as an integral 
operator with boundary at infinity: 
$$
\frac{{\mathds 1}}{\ri{\hat p}}\,g(x)=-\int_x^\infty\,g(t)\rd t . 
$$

The eigenfunctions of
${\hat H}$ are given by the Hurwitz zeta function $\psi_z(x) =-\zeta(z, x+1)$,
where $x\in(0,\infty)$ and $z=\half(1-\ri E)\in{\mathds C}$, and the minus sign
is our convention. The Hurwitz zeta function can be represented as a contour
integral
\begin{eqnarray}
-\zeta(z,x+1)=\frac{\Gamma(1-z)}{2\pi\ri}\int_C \frac{\re^{xt}t^{z-1}}
{1-\re^{-t}} \, \rd t, 
\label{e4}
\end{eqnarray} 
where the integration path $C$ is a Hankel contour that encircles the
negative-$t$ axis in the positive sense. For each $z$ the function $\psi_z(x)$
satisfies the relation 
$$
{\hat H} \psi_z(x)=\ri(2z-1) \psi_z(x).
$$
Letting $\phi_z(x)={\hat\Delta}\psi_z(x)$, a short calculation shows that
$\phi_z(x)=x^{-z}$, the eigenfunction of ${\hat h}_{\rm BK}$, and we have ${\hat
h}_{\rm BK}\phi_z(x)=\ri(2z-1)\phi_z(x)$.  

In \cite{r1} an inner product was introduced by making use of the biorthogonal 
systems associated with the Hamiltonian ${\hat H}$. It was then shown that 
${\hat H}$ is symmetric, that is, $\langle \varphi,{\hat H}\psi\rangle=\langle
{\hat H}\varphi,\psi\rangle$ with respect to the inner product, if the boundary 
condition $\psi(0)=0$ is satisfied. However, because $\psi_z(x)=-\zeta(z, x+1)$
and $\zeta(z,1)=\zeta(z)$ is the Riemann zeta function, the condition $\psi_z(0)
=0$ implies that $z$ must be a zero of $\zeta(z)$. The function $\zeta(z)$ has
both trivial and nontrivial zeros but the trivial zeros are excluded by imposing
a growth condition to ensure the orthogonality of the eigenfunctions with
respect to this inner product. It follows that if one can identify a suitable
space spanned by all eigenfunctions of ${\hat H}$ that satisfies the boundary
condition and a suitable growth condition, and if one can establish the 
essential selfadjointness of ${\hat H}$ on that space, one could conclude that
the Riemann hypothesis holds because $E=\ri(2z-1)$ is real only if the real part
of $z$ is $1/2$. 

\medskip
\noindent{\bf Relation to the Berry-Keating system}. As indicated above, ${\hat
H}$ is formally similar to the Berry-Keating Hamiltonian ${\hat h}_{\rm BK}$. In
fact, the two Hamiltonians are isospectral if the operators ${\hat\Delta}$ and
${\hat \Delta}^{-1}$ are both bounded \cite{r14}. That ${\hat p}$ is not
selfadjoint on ${\cal H}\equiv{\cal L}^2({\mathds R}^+,\rd x)$, with strictly
positive imaginary part, ensures the boundedness of these operators, and we find
that ${\hat H}$ and ${\hat h}_{\rm BK}$ are indeed isospectral on ${\cal H}$. 

It is known that ${\hat h}_{\rm BK}$ is (essentially) selfadjoint on ${\cal H}$.
The standard argument for showing this relies on the counting of von Neumann's 
deficiency indices \cite{r6,r7,r8}. More intuitively, one observes that ${\hat
h}_{\rm BK}$ generates an isometry in ${\cal H}$:
$$
\| \re^{-{\rm i}\lambda{\hat h}_{\rm BK}} f(x) \|_2=\left(\int_0^\infty\left|
\re^{-\lambda}f(\re^{-2\lambda}x)\right|^2\rd x\right)^{1/2}=\| f(x) \|_2,
$$
and then one applies Wigner's theorem. It has been shown in \cite{r8} that the 
eigenfunctions $\phi_z(x) = x^{-z}$ of ${\hat h}_{\rm BK}$, with eigenvalues 
$E=\ri(2z-1)\in{\mathds R}$, form singular basis in ${\cal H}$ in the sense that
$$
\int_0^\infty {\bar\phi}_{z'}(x) \phi_z(x)\, \rd x \propto \delta(E-E')
$$
and 
$$
\int_{-\infty}^\infty{\bar\phi}_{z}(x')\phi_z(x)\,\rd E\propto\delta(x-x'),
$$ 
even though $\{\phi_z(x)\}$ do not belong to ${\cal H}$. (This is analogous to
the momentum eigenstates in standard quantum mechanics. In fact, as shown 
in \cite{r6}, ${\hat h}_{\rm BK}$ is just the momentum operator $-\ri \rd/\rd q$ on 
${\mathds R}$ under the logarithmic map $x\to q=\ln x$.) The completeness of
$\{\phi_z(x)\}$ on ${\cal H}$, with $z=\half(1-\ri E)$ and $E$ real, then
implies that the Hurwitz zeta functions $-\zeta(z,x+1)$, with $z=\half(1-\ri E)$
and $E$ real, are complete on the space of functions $\{F(x)\}$ defined on $(-1,
\infty)$ such that $F(x)-F(x-1)\in{\cal H}$. 

As noted in \cite{r1}, the boundary condition $\psi_z(0)=0$ can be pushed
forward to the eigenspace of ${\hat h}_{\rm BK}$ by using ${\hat\Delta}$.
Specifically, by defining $f_z(x)\equiv\phi_z(x)-\zeta(z,x)$ we see that $\psi_z
(0)=0$ implies that $f_z(0)=0$. The vanishing of $f_z(x)$ at the origin is
possible only if $z=\half(1-\ri E)$ is a zero of the Riemann zeta function, so
we arrive at the quantization condition for the Berry-Keating Hamiltonian ${\hat
h}_{\rm BK}$ as a boundary condition. But the boundary condition imposes a
restriction of the Hilbert space ${\cal H}$, from which we conclude that $E_n$,
with $z_n=\half(1-\ri E_n)$ a nontrivial zero of $\zeta(z)$, is real, and the
set $\{E_n\}$ corresponds to the nontrivial zeros on the critical line. This,
however, does {\it not} establish the Riemann hypothesis because we have not
shown that the space of eigenfunctions $\phi_z(x)=x^{-z}$ of ${\hat h}_{\rm BK}$
that satisfy the boundary condition $f_z(0)=0$, but are not affiliated to ${\cal
H}$ (i.e. that do not belong to the basis states of ${\cal H}$), is empty. In
other words, the span of the totality of functions $\phi_z(x)=x^{-z}$ satisfying
the boundary condition $f_z(0)=0$ may be larger than ${\cal H}$, and unless one
can show that this is not the case, the selfadjointness of ${\hat h}_{\rm BK}$
is insufficient to arrive at the Riemann hypothesis. 

\medskip
\noindent{\bf Further comments on the quantization condition}. The boundary
condition $f_z(0)=0$ is not the only quantization condition for ${\hat h}_{\rm
BK}$. For instance, it has been suggested to the authors by Sarnak \cite{PS} and
Keating \cite{JK} that one may consider the space of functions $f(x)$ satisfying
the condition 
\begin{eqnarray}
\sum_{n=1}^\infty f(nx)=0, 
\label{eq:PS} 
\end{eqnarray}
where the sum is interpreted in a suitably regularized sense. If in particular 
$f(x)=\phi_z(x)$ is the eigenfunction of ${\hat h}_{\rm BK}$, then $f(nx)=n^{-z}
x^{-z}$ and hence (\ref{eq:PS}) implies that $z$ is a zero of the Riemann zeta
function. 

The quantization condition (\ref{eq:PS}) of Keating and Sarnak may be expressed 
alternatively in the operator formalism. Using the generator ${\hat x}{\hat p}=
({\hat h}_{\rm BK}+\ri)/2$ of dilations we have $f(nx)=n^{{\rm i}{\hat x}{\hat
p}} f(x)$, so (\ref{eq:PS}) can be expressed in the form 
$$\sum_{n=1}^\infty n^{{\rm i}{\hat x}{\hat p}}f(x)=0,$$ 
or more appropriately 
$$
\zeta\left(-{\rm i}{\hat x}{\hat p}\right)\,f(x)=0,
$$ 
to be interpreted in the sense of an analytic continuation of the summation 
representation. That is, for the eigenvalues of $-{\rm i}{\hat x}{\hat p}$ for 
which their real parts are greater than one, the action of $\zeta\left(-{\rm i}
{\hat x}{\hat p}\right)$ on $f(x)$ is defined by the left side of (\ref{eq:PS}),
in a way analogous to the idea presented in \cite{refLL}, but otherwise $\zeta
\left(-{\rm i}{\hat x}{\hat p}\right)$ can be interpreted as an integral
operator: 
$$
\zeta\left(-{\rm i}{\hat x}{\hat p}\right)\,f(x)\equiv\frac{1}{2\pi\ri}\int_{
u=0}^\infty\int_C\frac{t^{-1}\re^{-u}}{\re^{-t}-1}\,f\left(\frac{ux}{t}\right)
\rd t\,\rd u.
$$  

Thus, the space of functions satisfying (\ref{eq:PS}) consists of functions that
are annihilated by $\zeta(-{\rm i}{\hat x}{\hat p})$, the Riemann zeta function
valued at the dilation operator, which might appropriately be called the Riemann
dilation operator. Because $[{\hat h}_{\rm BK},{\hat x}{\hat p}]=0$ we have 
$$
\zeta\left(-{\rm i}{\hat x}{\hat p}\right) \, \phi_z(x) = \zeta(z) \, \phi_z(z) 
$$ 
for the Berry-Keating eigenfunctions $\phi_z(z)$. In short, the Riemann dilation
operator has the property that its eigenvalues are the Riemann zeta functions.
Any finite linear combination of $\phi_z(x)$ with all the $z$ belonging to the
set of zeros of $\zeta(z)$ is annihilated by $\zeta\left(-{\rm i}{\hat x}{\hat p
}\right)$. An open question is whether there are other functions annihilated by
$\zeta\left(-{\rm i}{\hat x}{\hat p}\right)$ \cite{PS}.

Because the quantization condition (\ref{eq:PS}) also leads to the zeros of the 
Riemann zeta function, it would be of interest to compare the lifting of
(\ref{eq:PS}) by using ${\hat\Delta}^{-1}$ with our boundary condition $\psi(0)=
0$. For the purpose of this comparison, let us first define the Kubert operator
${\hat T}_n$ according to the prescription 
$$
{\hat T}_n f(x)\equiv\sum_{k=0}^{n-1} f\left( \frac{x-k}{n}\right). 
$$
[Note that our convention of summing $f(x/n-k/n)$ differs from that used in 
\cite{K,KL} of summing $f(x/n+k/n)$.] Then, from 
$$
f\left(\frac{x}{n}\right)=n^{-{\rm i}{\hat x}{\hat p}}f(x)\qquad{\rm and}  
\qquad f\left(\frac{x-k}{n}\right)=\re^{-{\rm i}k{\hat p}}f\left(\frac{x}{n}
\right)
$$
we see that the Kubert operator can alternatively be expressed in the form 
\begin{eqnarray} 
{\hat T}_n=\frac{{\mathds 1}-\re^{-{\rm i}n{\hat p}}}{{\mathds 1}-\re^{-{\rm i
}{\hat p}}}\,n^{-{\rm i}{\hat x}{\hat p}}.
\label{eq:KO} 
\end{eqnarray}
In particular, from 
$${\hat T}_n\,\re^{tx}=\frac{1-\re^{-t}}{1-\re^{-t/n}}\,\re^{tx/n}$$
one sees at once that if $\psi_z(x)$ for $z\in{\mathds C}$ are the Hurwitz zeta 
functions (\ref{e4}), then for every $n$ we have 
\begin{eqnarray} 
{\hat T}_n\,\psi_z(x)=n^z\,\psi_z(x). 
\label{eq:xx5} 
\end{eqnarray} 
Conversely, the result of Milnor \cite{ref12} shows that $\psi_z(x)$ are the 
only functions that satisfy (\ref{eq:xx5}) for all $n$. 

With these preliminaries we consider the lifting of $f(nx)= n^{{\rm i}{\hat x}{
\hat p}}f(x)$ using ${\hat\Delta}^{-1}$. Commuting ${\hat\Delta}$ through ${\hat
x}{\hat p}$ a short calculation establishes that 
$${\hat\Delta}^{-1}f(nx)=n^{{\rm i}{\hat x}{\hat p}}\, \frac{{\mathds 1}-\re^{-{\rm
i}{\hat p}}}{{\mathds 1}-\re^{-{\rm i}n{\hat p}}}\,F(x),$$
where $F(x)={\hat\Delta}^{-1}f(x)$. Comparing this with the operator representation 
(\ref{eq:KO}) for ${\hat T}_n$ we thus deduce that 
\begin{eqnarray}
{\hat\Delta}^{-1}\sum_n f(nx)=\sum_n{\hat T}_n^{-1} \, F(x).  
\label{eq:QC} 
\end{eqnarray}
In particular, if $F(x)=\psi_z(x)$, then ${\hat T}_n^{-1}F(x)=n^{-z}F(x)$ for
all $n$, and we are left with the condition that $\zeta(z)=0$. In general,
however, the lifting of the Keating-Sarnak quantization condition, that is, the
vanishing of the right side of (\ref{eq:QC}), is a nonlocal condition, and thus
is different from our local boundary condition $F(0)=0$, even though both
conditions imply that $\zeta(z)=0$. 

\medskip
\noindent{\bf Fourier analysis.} Let us now turn to the momentum-space
representation of the differential equation eigenvalue problem. In \cite{r4}
Berry and Keating made an effective use of the Fourier analysis (on the whole
line ${\mathds R}$) to arrive at the phase factor associated with the Riemann
zeta function. We begin by taking the Fourier transform of the Hurwitz zeta
function $\psi_z(x)=-\zeta(z,x+1)$ on the half line ${\mathds R}^+$ to find that
\begin{eqnarray} 
{\hat\psi}_z(p)=\Gamma(1-z)\left[\frac{(-\ri p)^{z-1}}{1-\re^{{\rm i}p}}-\ri 
(2\pi)^{z-1}\left(\sum_{k=1}^\infty\frac{k^{z-1}}{p+2\pi k}-(-1)^{z} 
\sum_{k=1}^\infty\frac{k^{z-1}}{p-2\pi k}\right)\right]. 
\label{eq17}
\end{eqnarray} 
The local boundary condition $\psi_z(0)=0$ in $x$-space then translates into a
global condition in Fourier space ($p$ space) that the integral of ${\hat\psi
}_z(p)$ over $p$ must vanish. However, because ${\hat\psi}_z(p)$ is the
integrand in an integral representation for the Riemann zeta function  
$$\int_{{\rm i}\epsilon-\infty}^{{\rm i}\epsilon+\infty} {\hat\psi}_z(p)\,\rd p
=\zeta(z),$$
where $\epsilon>0$, the boundary condition forces $z$ to be a zero of $\zeta(z
)$. That ${\hat\psi}_z(x)$ is the integrand of an integral representation for
$\zeta(z)$ makes the Hamiltonian (\ref{e1}) special in analyzing specifically
the Riemann zeta function. However, the identification of the eigenvalue problem
in the momentum space in which (\ref{eq17}) is the solution remains an open
question. In what follows we sketch where the difficulties lie. 

Using the commutation relation $[{\hat x},{\hat p}]=\ri$, we reorder the
Hamiltonian (\ref{e1}) as
$${\hat H}=\frac{2{\hat p}}{{\mathds 1}-\re^{{\rm i}{\hat p}}}+2{\hat p}{\hat x}
+\ri{\mathds 1}.$$
We then consider the eigenvalue equation ${\hat H}g(p)=Eg(p)$ in momentum space.
By making the identification ${\hat x}=\ri\,\rd/\rd p$ we obtain the linear
first-order differential equation
\begin{eqnarray}
\left[\frac{2p}{1-\re^{{\rm i}p}}+2\ri p\frac{\rd}{\rd p}+\ri\right]g(p)=Eg(p).
\label{eq5}
\end{eqnarray}
Setting $z=\half(1-\ri E)$, an easy calculation shows that apart from a
multiplicative constant
\begin{eqnarray}
g(p)=(1-\re^{-{\rm i}p})^{-1} p^{z-1}.
\label{eq6}
\end{eqnarray}
While (\ref{eq6}) is the unique solution to (\ref{eq5}), its Fourier inversion 
does not quite agree with $\psi_z(x)=-\zeta(z, x+1)$. To see this, recall first
that because $x\in(0,\infty)$, the momentum operator ${\hat p}$ is not
selfadjoint and the momentum-space integration for Fourier inversion must be
performed along a straight line with a strictly positive
imaginary part in the complex-$p$ plane. (The Appendix gives a brief discussion
of Fourier analysis in the complex domain; see \cite{wiener} for details.) 

With this in mind, we examine 
$$f(x)=\frac{1}{2\pi}\int_{{\rm i}\epsilon-\infty}^{{\rm i}\epsilon+\infty} 
\re^{-{\rm i}px}\,g(p)\,\rd p,$$ 
where $\epsilon>0$. When $x\leq0$, closing the contour in the positive sense
gives no contribution. When $x>0$, closing the contour in the negative sense, we
find a branch cut emanating from the origin $p=0$ and poles at $p=2\pi n$, $n=
\pm1,\,\pm2,\,\ldots$. Hence for $x>0$ we get
$$f(x)=-\ri\sum_{\rm poles} \frac{1}{2\pi\ri}\oint\frac{\re^{-{\rm i}px} p^{z-1}
}{1-\re^{-{\rm i}p}}\rd p+\frac{\ri}{2\pi\ri}\int\limits_{\curvearrowright}
\frac{\re^{-{\rm i}px}p^{z-1}}{1-\re^{-{\rm i}p}}\rd p,$$
where $\oint$ denotes integration in the positive orientation. 
By changing the integration variable according to $-\ri p=t$, the contour
$\curvearrowright$ around the branch cut is rotated into $C$, so we obtain 
$$f(x)=-\sum_{n=-\infty,\neq0}^\infty(2\pi n)^{z-1} \re^{-2\pi{\rm i}nx} -\ri 
\frac{\ri^z}{2\pi\ri}\int_C \frac{\re^{tx} t^{z-1}}{1-\re^{t}} \rd t.$$

We examine the sum first:
\begin{eqnarray}
\sum_{n=-\infty,\neq0}^\infty \!\! \frac{\re^{-2\pi{\rm i}nx}}{(2\pi n)^{1-z}} 
&=& (2\pi)^{z-1} \left[ \sum_{n=1}^\infty \frac{\re^{-2\pi{\rm i}nx}}{n^{1-z}} 
+\sum_{n=-\infty}^{-1} \frac{\re^{-2\pi{\rm i}nx}}{n^{1-z}}\right]\nonumber\\
&=& (2\pi)^{z-1} \left[ \sum_{n=1}^{\infty} \frac{\re^{-2\pi{\rm i}nx}}{n^{1-z}}
-\frac{1}{(-1)^{-z}}\sum_{n=1}^\infty \frac{\re^{2\pi{\rm i}nx}}{n^{1-z}}\right]
\nonumber \\
&=& (2\pi)^{z-1} \left[\sum_{n=1}^{\infty} \frac{\re^{-2\pi{\rm i}nx}}{n^{1-z}}
- \re^{\pi{\rm i} z}\sum_{n=1}^\infty \frac{\re^{2\pi{\rm i}nx}}{n^{1-z}}\right]
\nonumber \\
&=&-(2\pi)^{z-1} \re^{\pi{\rm i} z/2} \left[ \re^{\pi{\rm i} z/2}\sum_{n=1
}^\infty \frac{\re^{2\pi{\rm i}nx}}{n^{1-z}} - \re^{-\pi{\rm i}z/2}
\sum_{n=1}^{\infty} \frac{\re^{-2\pi{\rm i}nx}}{n^{1-z}}\right]. 
\nonumber 
\end{eqnarray}
This result calls to mind an identity for the Hurwitz zeta function \cite{ref3}
$$\frac{1}{\Gamma(1-z)}\,\zeta(z,x)=-\ri\,(2\pi)^{z-1}\left(\re^{\pi{\rm i}
z/2}\sum_{n=1}^{\infty}\frac{\re^{2\pi{\rm i}nx}}{n^{1-z}}-\re^{-\pi{\rm i}z/2}
\sum_{n=1}^\infty\frac{\re^{-2\pi{\rm i}nx}}{n^{1-z}}\right),$$
which is valid for $0<x\leq1$ and for $\Re(z)<0$. From this identity it follows
that
$$
-\sum_{n=-\infty,\neq0}^\infty \!\! (2\pi n)^{z-1} \re^{-2\pi{\rm i}nx} = 
\frac{\ri\,\re^{\pi{\rm i}z/2}}{\Gamma(1-z)}\, \zeta(z,x) .
$$ 
As for the integral term, we get 
$$
-\ri \frac{\ri^z}{2\pi\ri}\int_C\frac{\re^{tx} t^{z-1}}{1-\re^{t}} \, \rd t 
= - \frac{\ri \,\re^{\pi{\rm i}z/2}}{\Gamma(1-z)}\, \zeta(z,x) .  
$$ 
Combining these results, we deduce that  
$$
f(x)=0 
$$
for $0<x\leq1$. Apart from the fact that the summation above was taken outside
the range of values of $z$ of interest, we are led to conclude that $f(x)\neq
\psi_z(x)$.

There are several reasons why this naive analysis gives an incorrect expression.
First, we consider taking the Fourier transform of the infinite-order
differential equation ${\hat H}\psi(x)=E\psi(x)$ in $x$-space. Specifically, by
expanding ${\hat H}$ as a series in powers of $\ri{\hat p}=\rd/\rd x$ and taking
into account the generating function  
\begin{eqnarray}
\frac{t}{\re^t-1}=\sum_{n=0}^\infty B_n\frac{t^n}{n!}
\label{BN} 
\end{eqnarray}
for the Bernoulli numbers, the differential equation takes the form 
\begin{eqnarray} 
\ri\left[ 2\sum_{k=0}^\infty\frac{B_k}{k!} \left(\frac{\rd}{\rd x}\right)^k 
-2x\frac{\rd}{\rd x}-1\right]\psi(x)=E\psi(x) . 
\label{eq16}
\end{eqnarray} 
Multiplying both sides by $\re^{{\rm i}px}$ and integrating with respect to $x$,
the usual procedure is to integrate by parts to replace each differential
operator $\rd/ \rd x$ by $\ri p$, assuming that such an integration is
permissible. However, unlike a typical problem on the real line, the present
case involves a half line. Thus, although we have the boundary condition $\psi(
0)=0$, the derivatives of $\psi(x)$ do not vanish at the origin. Consequently,
we are left with a differential equation of the form (\ref{eq5}), along with
infinitely many boundary terms involving the derivatives of the Riemann zeta 
function evaluated at the Riemann zeros. Second, the radius of convergence of 
the expansion (\ref{BN}) is $2\pi$, which means that term-by-term integration of
(\ref{eq16}) in $x$ over the range $(0,\infty)$ need not be valid unless it can
be justified.

These issues do not arise in the Fourier analysis of Berry and Keating because 
the eigenfunctions of ${\hat h}_{\rm BK}$ are considered on the whole line in 
\cite{r4}, and because the differential operator is first order in both position
and momentum space. We therefore observe that the identification of the 
momentum-space representation of our differential-equation eigenvalue problem 
to which (\ref{eq17}) is the solution remains an interesting open problem. 
Here, instead, to proceed further with the analysis of the Hamiltonian
(\ref{e1}), we consider applying WKB analysis to the differential equation
(\ref{eq16}). 

\medskip
\noindent{\bf WKB analysis.} In the foregoing analysis we have chosen units such
that Planck's constant takes the value $\hbar=1$. However, for the purpose of a
WKB analysis we reinstate $\hbar$ so that we have $\ri{\hat p}=\hbar\,\rd/\rd x$
and the differential equation (\ref{eq16}) becomes 
\begin{eqnarray} 
\ri\hbar\left[ 2\sum_{k=0}^\infty \hbar^k \frac{B_k}{k!}\left(\frac{\rd}{\rd x}
\right)^k -2x\frac{\rd}{\rd x}-1\right] \psi(x) = E \psi(x) . 
\label{eq:z1}
\end{eqnarray} 
Following the approach of WKB theory, we make the usual {\it ansatz} for $\psi(x
)$:
\begin{eqnarray}
\psi(x)=\exp\left( \frac{1}{\hbar} \sum_{n=0}^\infty S_n(x) \hbar^n \right)  
\label{eq:z2} 
\end{eqnarray} 
and substitute (\ref{eq:z2}) into (\ref{eq:z1}). Writing $\psi^{(k)}(x)=\rd^k
\psi(x)/\rd x^k$, we express (\ref{eq:z1}) more succinctly as 
\begin{eqnarray}
2\sum_{k=0}^\infty \hbar^k \frac{B_k}{k!} \frac{\psi^{(k)}(x)}{\psi(x)} - 2x 
\frac{\psi^{(1)}(x)}{\psi(x)} = 1 + \frac{E}{\ri\hbar} . 
\label{eq:21} 
\end{eqnarray}

By equating terms of order $\hbar^{-1}$ we obtain 
\begin{eqnarray}
-2 \ri x S_0'(x)=E,
\label{eq:z3} 
\end{eqnarray} 
where $S_0'(x)=\rd S_0(x)/\rd x$. This can readily be solved to yield 
\begin{eqnarray}
S_0(x)=\frac{\ri E}{2}\,\ln x. 
\label{eq:23} 
\end{eqnarray}
Thus, setting $\hbar=1$ the leading-order WKB ansatz gives the result $\psi(x)
\sim x^{{\rm i}E/2}$. This result is interesting because the leading-order WKB
approximation gives the {\it eikonal} (geometric optics) approximation to the
oscillatory phase behavior, and this requires that $E$ be real. The discovery
that $\psi(x)\sim x^{{\rm i}E/2}$ is therefore consistent with our hypothesis
that the eigenvalues of ${\hat H}$ are real in a Hilbert space to which all its 
eigenstates satisfying the boundary condition $\psi(0)=0$ are affiliated. 

Next, we consider terms of order $\hbar^0$ in (\ref{eq:21}) and obtain 
$$-1-2xS_1'(x)+2\sum_{n=0}^\infty\frac{B_k}{k!}\big(S_0'(x)\big)^k=0.$$
By using (\ref{BN}) and (\ref{eq:z3}), we deduce that 
$$S_1'(x)=-\frac{1}{2x}+\frac{\ri E}{2x^2(\re^{{\rm i}E/2x}-1)}.$$
This can be integrated to yield the simple expression 
\begin{eqnarray}
S_1(x)=-\frac{1}{2}\,\ln x-\ln\left(1-\re^{-{\rm i}E/2x}\right).
\label{eq:26} 
\end{eqnarray} 
The terms of order $\hbar^0$ in a WKB analysis provide the {\it physical-optics}
approximation, which determines the leading asymptotic behavior of the amplitude
of $\psi(x)$ for large $x$. Combining (\ref{eq:23}) and (\ref{eq:26}) and
setting $\hbar=1$, we find that for $x\gg1$, $\psi(x)\sim x^{\frac{1}{2}(1+{\rm
i}E)}$. Remarkably, this is in exact agreement with the asymptotic behavior of
the Hurwitz zeta function $\zeta(z, x+1)$ obtained in \cite{r1} that $\zeta(z,x
+1)\sim x^{1-z}= x^{\frac{1}{2}(1+{\rm i}E)}$. 

Including higher-order terms in a WKB series typically does not affect the
leading asymptotic behavior. However, as a consistency check, in the following
we calculate the terms of order $\hbar$ explicitly to verify that the leading
asymptotic behavior of $\psi(x)\sim x^{\frac{1}{2}(1+{\rm i}E)}$ is unchanged.
Equating coefficients of $\hbar$ in (\ref{eq:21}), we find that
\begin{eqnarray}
S_2'(x)=\frac{1}{x}\sum_{n=0}^\infty\frac{B_n}{n!}\left[n\big(S_0'(x)
\big)^{n-1}S_1'(x)+\frac{n(n-1)}{2}\big(S_0'(x)\big)^{n-2}S_0''(x)\right].
\label{eq:z4} 
\end{eqnarray} 
Substituting the expressions for $S_0'(x)$, $S_0''(x)$, and $S_1'(x)$ in
(\ref{eq:z4}) and using the relations 
$$\sum_{k=1}^\infty \frac{B_k}{(k-1)!}t^{k-1}=\frac{1}{\re^t-1}-\frac{t\re^t}
{(\re^t-1)^2}$$
and 
$$\sum_{k=2}^\infty\frac{B_k}{(k-2)!}t^{k-2}=-\frac{(t+2)\re^t}{(\re^t-1)^2}+
\frac{2t\re^{2t}}{(\re^t-1)^3},$$
we deduce that 
\begin{eqnarray}
S_2'(x) &=& \frac{1}{2x} \left( -\frac{1}{x} + \frac{\ri E}{x^2(\re^{{\rm i}E/
2x}-1)}\right)\left(\frac{1}{\re^{{\rm i}E/2x}-1}-\frac{\ri E\,
\re^{{\rm i}E/2x}}{2x(\re^{{\rm i}E/2x}-1)^2}\right)\nonumber\\
& & +\frac{\ri E}{4x^3} \left[\frac{2\,\re^{{\rm i}E/2x}}{(\re^{{\rm i}E/2x}
-1)^2}+\frac{\ri E\,\re^{{\rm i}E/2x}}{2x(\re^{{\rm i}E/2x}-1)^2}-
\frac{\ri E\,\re^{{\rm i}E/x}}{x(\re^{{\rm i}E/2x}-1)^3}\right].
\label{eq:z5} 
\end{eqnarray} 
After some lengthy algebra we find that remarkably (\ref{eq:z5}) can be
integrated in closed form to yield 
$$S_2(x)=\frac{1}{2x\big(\re^{{\rm i}E/2x}-1\big)}-
\frac{\ri E (1+\re^{{\rm i}E/2x})}{4x^2\big( 1-\re^{{\rm i}E/2x}\big)^2}.$$
We then find that for $x\gg1$, 
$$\re^{S_2(x)}\sim\re^{7{\rm i}/E}\left[1+\frac{7}{4x}+\frac{49-10\ri E}{32x^2}
\right],$$
and thus, as anticipated, $S_2(x)$ does not affect the leading asymptotic
behavior of $\psi(x)$. 

\medskip
\noindent{\bf Discussion.} After a close examination of the quantization
condition for the Hamiltonian ${\hat H}$ defined in (\ref{e1}), we considered
(a) Fourier analysis, and (b) WKB analysis, of the differential-equation
eigenvalue problem ${\hat H}\psi(x)=E\psi(x)$ associated with ${\hat H}$. A
naive Fourier inversion of the eigenvalue problem gives a result that is not the
Fourier transform (\ref{eq17}) of the correct solution, leading to the open
problem of identifying the differential equation eigenvalue problem in Fourier
space. However, using WKB analysis, we have performed an elaborate calculation
to find in closed form the leading three terms in the approximation. We remark
that WKB analysis of an {\it infinite-order} differential equation is
unconventional; ordinarily, WKB is used to approximate solutions to second-order
differential equations (see \cite{ref5.0,ref5} for exceptions). Our result for
the asymptotic behavior of the eigenfunction is shown to be in exact agreement
with the asymptotic behavior of the eigenfunction worked out in \cite{r1}.
Leading-order WKB analysis also suggests that the eigenvalues of ${\hat H}$ are
real. Several open problems remain: (i) What is the space of all functions
annihilated by the Riemann dilation operator $\zeta(-\ri{\hat x}{\hat p})$? (ii)
What is the differential-equation eigenvalue problem to which (\ref{eq17}) is
the solution? (iii) Can one show that the span of the eigenfunctions of 
${\hat h}_{\rm BK}$ satisfying the boundary condition $f_z(0)=0$ is not larger 
than ${\cal H}$? 

\medskip
\noindent{\bf Appendix: Generalized Fourier transforms.} Assume that the
function $f(x)$, $x>0$, is such that $\re^{-\alpha x}f(x)\to0$ as $x\to+\infty$
for a given $\alpha>0$. For such a function $f(x)$ we define 
$$F_+(w)\equiv\int_0^\infty f(x)\,\re^{{\rm i}xw}\,\rd x.$$
If we write $w=u+\ri v$, then 
$$F_+(w)=\int_0^\infty f(x)\,\re^{{\rm i}xu}\,\re^{-xv}\,\rd x$$
exists if $v>\alpha$. Then in the complex-$w$ plane (i) $F_+(w)$ is analytic in 
the region $\Im(w)>\alpha$, and (ii) $F_+(w)\to0$ as $v\to\infty$. 

We now define 
$$g(x)\equiv\left\{\begin{array}{cc}f(x)\,\re^{-xv} & ~(x>0),\\ 0 &~(x\leq0).
\end{array} \right.$$
Then $F_+(w)$ is an ordinary Fourier transform,
$$\int_{-\infty}^\infty g(x)\,\re^{{\rm i}xu}\,\rd x=F_+(u+\ri v),$$
and
$$g(x)=\frac{1}{2\pi}\int_{-\infty}^\infty F_+(u+\ri v)\,\re^{-{\rm i}ux}\,\rd
u.$$
In other words, 
\begin{eqnarray}
\frac{1}{2\pi}\int_{-\infty}^\infty F_+(u+\ri v)\,\re^{-{\rm i}(u+{\rm i}v)x}\,
\rd u &=& \frac{1}{2\pi} \int_{{\rm i}\alpha-\infty}^{{\rm i}\alpha+\infty} 
F_+(w)\,\re^{-{\rm i}wx}\,\rd w\nonumber\\
&=& \frac{1}{2\pi}\oint F_+(w)\,\re^{-{\rm i}wx}\,\rd w\nonumber\\
&=& \left\{ \begin{array}{cc}f(x)& ~(x>0),\\ 0 & ~(x\leq0).\end{array}\right. 
\nonumber 
\end{eqnarray} 

\noindent 
{\bf Acknowledgement}.
DCB thanks the Russian Science Foundation for support (project 16-11-10218). We
thank J. Keating and P. Sarnak for stimulating discussions.


\begin{thebibliography}{999}

\bibitem{r1} C.~M.~Bender, D.~C.~Brody, and M.~P.~M\"uller (2017) 
Hamiltonian for the zeros of the Riemann zeta function. 
{\it Physical Review Letters} {\bf 118}, 130201. 

\bibitem{r4}
M.~V.~Berry and J.~P.~Keating (1999) 
H=xp and the Riemann zeros. 
In {\em Supersymmetry and Trace Formulae: Chaos and Disorder}. 
Edited by I.~V.~Lerner {\em et al}. (New York: Kluwer Academic / Plenum)

\bibitem{refxx} 
N. I. Akhiezer and I. M. Glazman (1961) 
{\em Theory of Linear Operators in Hilbert Space} (New York: Ungar).

\bibitem{r14} D.~C.~Brody, 
Biorthogonal quantum mechanics. 
{\em J. Phys. A: Math. Theor.}~{\bf 47}, 035305 (2014). 

\bibitem{r6} J.~Twamley and G.~J.~Milburn (2006) 
The quantum Mellin transform. 
{\em New Journal of Physics} {\bf 8}, 328.

\bibitem{r7} G.~Sierra (2007) 
$H = xp$ with interaction and the Riemann zeros. 
{\em Nuclear Physics} B {\bf 776}, 327.

\bibitem{r8} S.~Endres and F.~Steiner (2010) 
The Berry-Keating operator on $L^2({\mathds R}_>, \rd x)$ and on
compact quantum graphs with general self-adjoint realizations. 
{\em Journal of Physics} A {\bf 43}, 095204. 

\bibitem{PS} 
P.~Sarnak (2017) Private communication. 

\bibitem{JK} 
J.~P.~Keating (2018) Private communication. 

\bibitem{refLL} 
J.~C.~Lagarias and W.-C.~W.~Li (2016) 
The Lerch zeta function IV. Hecke operators. 
{\em Res. Math. Sci.} {\bf 3}:33. 

\bibitem{K}
D.~Kubert (1979) 
The universal ordinary distribution. 
{\em Bull. Math. Soc. Fr.} {\bf 101}, 179--202. 

\bibitem{KL} 
D.~Kubert and S.~Lang (1981) 
Modular Units. 
{\em Grundlehren Wissenschaften} {\bf 244}. Berlin: Springer. 

\bibitem{ref12} 
J.~Milnor (1983) 
On polylogarithms, Hurwitz zeta functions, and the Kubert identities. 
{\em L'Enseignement Math\'ematique}, {\bf 29}, 281-322.

\bibitem{wiener} N.~Wiener and R.~C.~Paley (1934)  
{\em Fourier Transforms in the Complex Domain}. 
(American Mathematical Society, Providence, RI).

\bibitem{ref3} 
T.~Arakawa, T.~Ibukiyama, and M.~Kaneko (2014) 
{\em Bernoulli Numbers and Zeta Functions}. 
(Springer Monographs in Mathematics, Springer, Tokyo). 

\bibitem{ref5.0} 
R.~B.~Dingle and G.~J.~Morgan (1968) 
WKB methods for difference equations II. 
{\it  Applied Scientific Research} {\bf 18}, 238. 

\bibitem{ref5} 
C.~M.~Bender, F.~Cooper, G.~S.~Guralnik, D.~H.~Sharp, and M.~L.~Silverstein
(1979) Multilegged propagators in strong-coupling expansions. 
{\it Physical Review} D {\bf 20}, 1374. 

\end{thebibliography}
\end{document}